# All dielectric integrable optical isolators

SEVAG ABADIAN, GETÚLIO SOUZA, STANISLAV WINKLER, MARIAN BOGDAN SIRBU, MICHAIL SYMEONIDIS, TOLGA TEKIN

*Fraunhofer-Institut für Zuverlässigkeit und Mikrointegration (IZM), Gustav-Meyer-Allee 25, 13355 Berlin, Germany*
*sevag.abadian@izm.fraunhofer.de*

**Abstract:** On-chip optical isolators, functioning as unidirectional gates for light, play a crucial role in maintaining signal integrity, preventing laser destabilization, and fortifying the overall performance of optical systems. In this paper, we propose a five-layered heterostructure consisting of a magneto-optic material sandwiched between parallel dielectric slab waveguides. Under TMOKE configuration, the coupled optical modes undergo an electromagnetic profile transformation that can be harnessed to confine the input mode in one waveguide during forward propagation and in the other during backward propagation. Together with a ring resonator, such a system can provide a 20dB isolation ratio with negligible insertion losses.

### 1. Introduction

The surging reliance on online services, streaming platforms, and the pervasive integration of Internet of Things (IoT) devices into our daily lives are steadily amplifying the workload on data centers, underscoring the critical need for hardware enhanced efficiency. Photonic Integrated Circuits (PICs) play an essential role fulfilling the requirement for fast data processing with minimal power consumption. The effectiveness of these circuits is dependent upon the seamless integration of the different photonic components capable of optimizing signal propagation, minimizing power consumption, and ensuring reliable data transmission. Among these critical components, integrable routing elements such as isolators, switches, and circulators play a significant role in optimizing circuit functionality. Isolators, specifically, are crucial for stabilizing laser diode emissions, contributing to the overall stability and performance of these circuits. Achieving low losses and efficient functionality in these routing components is key. The quest of on-chip integrable optical isolators with reduced footprint and low power consumption has been ongoing for decades. Researchers have explored various architectural designs, each utilizing different mechanisms to achieve time and spatial symmetry breaking.

Over the past decades, these components have been demonstrated both experimentally and theoretically. Various mechanisms with different architectures have been proposed and thoroughly explored based on magneto-optics (MO) [1], electro-optics (EO) [2], acousto-optics (AO) [3], optomechanical (OM) [4], photonic transition (PT) [5], Photonic crystal (PhC) [6], and piezoelectric (PZT) interactions [7].

For isolators, one of the first attempts was to reproduce the principle of the bulk Faraday isolators. Low forward and high backward insertion loss (*IL*) are the characteristics of this bulky isolator which is incompatible with PICs. This isolator requires the integration of polarizers into the system that has yet to be accomplished. From here, there is a push towards achieving isolators in waveguiding configuration [8].

Among the different MO effects, the Transverse Magneto-Optical Kerr Effect (TMOKE) appears as the most compatible solution due to its advantage of not effecting the polarization of the input light.

In the Non-Reciprocal Losses (NRL), the combination of a semiconductor optical amplifier (SOA) with a ferromagnetic coating is investigated. By injecting a current into the SOAs the losses in the forward direction can be compensated. The polarization-dependent design configures the imaginary component of the effective refractive index, resulting in unequal losses in the opposite propagation directions. The design was validated in TM configuration [9] at $\lambda_0 = 1.3\ \mu m$, and in TE configuration [10] at a $\lambda_0 = 1.55\ \mu m$, which demonstrated approximately $IR = 99\ dB/cm$ and $IR = 14.7\ dB/mm$, respectively.

In the Non-Reciprocal Phase Shift (NRPS) effect, a change in the real part of the effective refractive index is observed, which results in a phase difference between the forward and backward signals. Here, an MZI is covered with a magnetic rare garnet, which means that, light would constructively interfere in the forward sense and destructively in the backward sense. Many research groups exploited the NRPS; an $IR = 19\ dB$ was achieved at $\lambda_0 = 1.54\ \mu m$ with a $L = 8\ mm$ [11].

In resonator devices, many groups developed highly resonant microrings. Approaches varied with respect to the geometry (radius) and placement (filling material: in the disk, cladding material: applied on top of the resonator). Regarding the first scenario, $IR = 20\ dB$ was recorded, along with $IL$ lower than $0.1\ dB$ and a bandwidth of $0.4\ nm$ [12]. As for the cladding solution, a first demonstrator exhibited $IR = 19.5\ dB$ for $L = 290\ \mu m$ [13], while a second achieved $9\ dB\ IR$ at $\lambda_0 = 1.55\ \mu m$ [14].

Magnetoplasmonic isolators based on the TMOKE effect were also investigated. In this case, garnets and metals were integrated to leverage the SPP confinement along the metal/dielectric interface. In the design of magnetoplasmonic MZIs, researchers achieved an $IR = 22.82\ dB$ with low $IL$ [15]. Others utilized a magnetoplasmonic slot guide to excite an LRSPP mode using a taper. This resulted in an $IR = 30\ dB$, at $\lambda_0 = 1.55\ \mu m$ [16], although promising results were shown for a very compact structure, however for a complete device with input/output couplers and adiabatic tapers, the isolator would have $IL > 10\ dB$.

In this paper, we contribute to this field by exploring dielectric-based MO isolators. Our work aims to add to the existing knowledge surrounding efficient isolator design and integration within PICs, presenting a novel approach that holds promise for enhancing both speed and energy efficiency in data center operations.

We start by presenting the principle underscoring the proposed waveguiding isolator. Using Maxwell's equations, we derive the wave equations required for coupled even/odd mode analysis. This step is crucial to understand the properties of the hosted optical modes by the five layered heterostructure. Coupled Mode Theory (CMT) is introduced to accurately capture the propagation and interaction of the coupled modes while traversing the parallel slab waveguides. Moreover, the additional element that offers the non-reciprocal service in the optical isolator is discussed. To conclude, the complete design is simulated and the characteristics of the isolator presented.

## 2. Waveguide design

The waveguiding architecture exploited here is referred to as the five-layered heterostructure, where slab waveguides a and b (referred to as *Wa* and *Wb*, respectively) are placed in parallel, surrounding a MO material and placed in cladding.

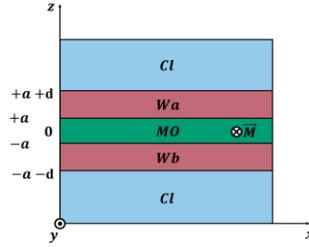

Fig. 1. Top view of the five-layered heterostructure. The MO material is placed between parallel waveguides, all inserted within the cladding. The magnetic field, $\vec{M}$, is applied perpendicular to the electric field component in the propagation direction.

Although different configurations can provide similar performance, in this paper, we focus on the theoretical derivations and modeling of electromagnetic fields along with their coupled propagation for the one presented in Fig. 1. In the absence of MO materials, the parallel dielectric waveguides support coupled optical even and odd transverse magnetic (TM) modes that display symmetry and antisymmetry in their magnetic field profiles around the central axis. While propagating in parallel waveguides, these modes do not exchange power. However, with the introduction of MO material under an external magnetic field, specifically in the TMOKE configuration, these modes become asymmetric and anti-asymmetric.

When the external magnetic field is applied inwards, in the forward direction, if the even mode is concentrated in one waveguide, then in the other direction the even mode will be concentrated in the second waveguide (Fig. 2a, b). It is noteworthy that a flipping of the external magnetization direction from inwards to outwards results in a corresponding inversion of this behavior (Fig. 2c, d).

To mitigate coupled mode interaction and regulate power exchange within the parallel region, each waveguide is strategically connected to S-bends on both the left and right sides, facilitating the excitation of the desired modes in both senses.

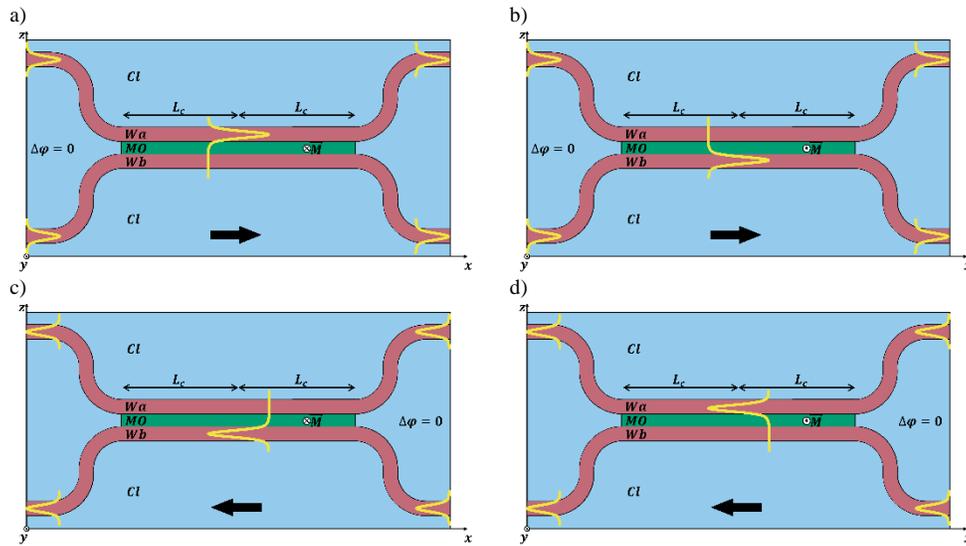

Fig. 2. Top view of the five layered heterostructure connected with S-bends excited by injecting modes with $\Delta\varphi = 0$ to excite the even mode (represented in yellow bold). Black bold arrows represent the propagation direction. Forward propagation for external magnetic field a) inwards and b) outwards directions, respectively. Backward propagation for external magnetic field c) inwards and d) outwards directions, respectively.

Hence, inverting the direction of propagation from forward to backward is equivalent to changing the direction of the externally applied magnetic field from inwards to outwards. As it was shown in Fig. 2, for the even mode to be excited, the phase difference is set to zero, i.e. $\Delta\varphi = 0$. Before delving further into the underlying physical effects, it is pertinent to highlight also the phase difference between the excited modes at the two input ports.

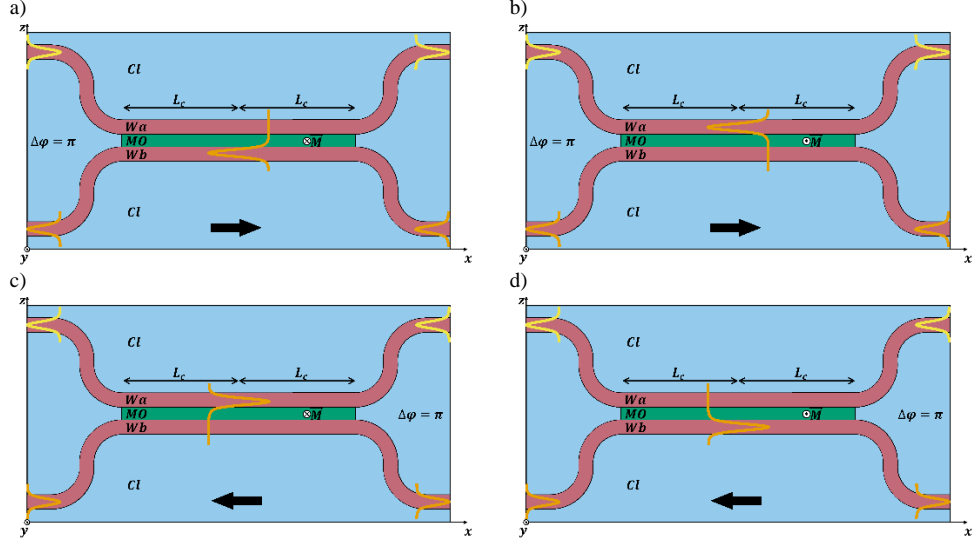

Fig. 3. Top view of the five layered heterostructure connected with S-bends excited by injecting modes with $\Delta\varphi = \pi$ to excite the odd mode (represented in orange bold). Black bold arrows represent the propagation direction. Forward propagation for external magnetic field a) inwards and b) outwards directions, respectively. Backward propagation for external magnetic field c) inwards and d) outwards directions, respectively.

In Fig. 3, upon exciting the input ports, the modes converge at the parallel region, initiating coupling. Upon propagating a specified distance, all input power converges into one of the two waveguides. For $\Delta\varphi = \pi$, the odd mode emerges in waveguide $Wb$ in the forward direction with an inward magnetic field (Fig. 3a) and in the backward direction with an outward magnetic field (Fig. 3d). Conversely, in waveguide $Wa$, the odd mode appears in the forward direction with an outward magnetic field (Fig. 3b) and in the backward direction with an inward magnetic field (Fig. 3c).

In the proposed heterostructure, control over light path ($Wa$ or $Wb$) is achievable through three distinct factors: the direction of light propagation, the orientation of the externally applied magnetic field, and the phase difference between the two injected modes at the input ports.

## 3. Modal analysis

In the quest to break time symmetry, employing transverse MO materials proves to be an efficient strategy. The dielectric tensor of a MO material can be represented in the form [17]:

$$\varepsilon_d = \begin{pmatrix} \varepsilon_{xx} & \varepsilon_{xy} & \varepsilon_{xz} \\ \varepsilon_{yx} & \varepsilon_{yy} & \varepsilon_{yz} \\ \varepsilon_{zx} & \varepsilon_{zy} & \varepsilon_{zz} \end{pmatrix} \quad (1)$$

Each pair of these non-diagonal elements are equal and opposite in magnitude ($\varepsilon_{xy} = -\varepsilon_{yx}$, $\varepsilon_{xz} = -\varepsilon_{zx}$, $\varepsilon_{yz} = -\varepsilon_{zy}$). In the absence of an external magnetic field, this tensor could be written as a diagonal one where the three elements are equal to each other ($\varepsilon_{xx} = \varepsilon_{yy} = \varepsilon_{zz}$).

The off-diagonal element is given by Eq. (2).

$$\varepsilon_{xy} = i \cdot g = i \cdot n_{MO} \cdot \lambda \cdot \theta_F / \pi \quad (2)$$

Where: $g$ is the MO gyrotropic parameter, $\varepsilon_d$ is the dielectric constant, $\lambda$ is the wavelength of operation and $\theta_F$ is the associated Faraday rotation coefficient.

For the sake of simplicity in theoretical derivations, analysis, and numerical simulations, and to maintain generality without compromising rigor, we consider our system as a two-dimensional structure—devoid of field variations in the y direction. It is essential to note that, in our analysis, we maintain the material permittivities as independent of frequency. This choice does not impose any limitations on our conceptual framework. If necessary, the dispersion of material parameters can be readily incorporated into the initial design of the system.

Using Maxwell's equations, we derive the generalized dispersion relations. A schematic of the five layered heterostructure is shown in Figure 1a. The design shows a MO material with width $2a$ and dielectric $Wa$ and $Wb$ of width $d$. This type of structure can host the even and odd coupled modes. These eigenmodes arise from evanescent coupling between the optical modes supported by the individual $Wa$ and $Wb$ which propagate in the x direction.

In a cartesian coordinate system the electromagnetic fields can be represented as:

$$\vec{E} = \vec{E_0} e^{i(\omega t - \beta x)} \qquad (3)$$
$$\vec{H} = \vec{H_0} e^{i(\omega t - \beta x)} \qquad (4)$$

Where $\vec{E_0}$ and $\vec{H_0}$ are the amplitude vectors with complex-valued components. $\beta$ is the propagation constant of the traveling waves and corresponds to the component of the wave vector in the direction of propagation and $\omega = 2\pi f$ is the angular frequency of the incident wave.

By selecting appropriate modal field functions for each region of the waveguide and applying electromagnetic field boundary conditions, one can derive the modal field solution and the characteristic equation. The magnetic field components for the structure in Fig. 1., can be expressed as follows:

$$H_y(z) = \begin{cases} A_1 \exp[-K_1(z - a - d)] & z \geq a + d \\ A_2 \cos[K_2(z - a - d/2)] & a \leq z \leq a + d \\ A_3 \exp[K_3 z] + A_4 \exp[-K_3 z] & -a \leq z \leq a \\ A_5 \cos[K_4(z + a + d/2)] & -a - d \leq z \leq -a \\ A_6 \exp[K_5(z + a + d)] & z \leq -a - d \end{cases} \qquad (5)$$

By replacing $H_y$ of each layer into the wave equation, one can find the expression for the wavevectors in each layer of the considered structure (Eq. (6)).

$$K_{1,3,5} = \sqrt{(\beta_{e,o}^2 \gamma_{xx}^{1,3,5} - k_0^2)/\gamma_{xx}^{1,3,5}}, \quad K_{2,4} = \sqrt{(k_0^2 - \beta_{e,o}^2 \gamma_{xx}^{2,4})/\gamma_{xx}^{2,4}} \qquad (6)$$

Where:

$$\gamma_{zx}^3 = ig/(n_3^4 - g^2), \quad \gamma_{xx}^3 = n_3^2/(n_3^4 - g^2)$$
$$\gamma_{xx}^{1,5} = 1/n_{1,5}^2, \quad \gamma_{xx}^{2,4} = 1/n_{2,4}^2 \qquad (7)$$

Finally, applying the continuity of the electromagnetic fields $E_x$, $H_y$ and $E_z$ at the different boundaries $(-a - d, -a, +a, +a + d)$, for this wave, the complex transcendental relation is written in the following way:

$$K_2 d = \tan^{-1}\left(\frac{\gamma_{xx}^1 K_1}{\gamma_{xx}^2 K_2}\right) + \tan^{-1}\left(\frac{A_3 \exp[K_3 a](\beta_{e,o} \gamma_{zx}^3 - i\gamma_{xx}^3 K_3) + A_4 \exp[-K_3 a](\beta_{e,o} \gamma_{zx}^3 + i\gamma_{xx}^3 K_3)}{-i\gamma_{xx}^2 K_2 (A_3 \exp[K_3 a] + A_4 \exp[-K_3 a])}\right) \qquad (8)$$

Where:

$$A_3 = \frac{(-2\beta\gamma_{zx}^3\cosh(2K_3a)A_4 \pm \sqrt{(2\beta_{e,o}\gamma_{zx}^3\cosh(2K_3a)A_4)^2 - 4(\beta_{e,o}\gamma_{zx}^3 - i\gamma_{xx}^3 K_3)(\beta_{e,o}\gamma_{zx}^3 + i\gamma_{xx}^3 K_3)A_4^2}}{2(\beta_{e,o}\gamma_{zx}^3 - i\gamma_{xx}^3 K_3)} \quad (9)$$

In Eq. (9), the two solutions obtained from the opposite signs correspond to the propagation constants of the even and odd modes, denoted as $\beta_e$ and $\beta_o$, respectively. Notably, the relation between $A_3$ and $A_4$ reveals distinct amplitudes of the $H_y(z)$ component for each of the coupled modes, $A_2$ and $A_5$ in the waveguides *Wa* and *Wb*, respectively. This disparity signifies that the even and odd coupled modes deviate from their conventional symmetric and anti-symmetric behaviors, adopting asymmetric and anti-asymmetric characteristics around the z-axis. By solving the left and right parts of the dispersion relation in Eq. (8) for the two cases of $A_3$, one can determine the propagation constants of the even and odd coupled modes.

Upon increasing the width of the MO layer, the system can be effectively partitioned into two sections (Fig. 4).

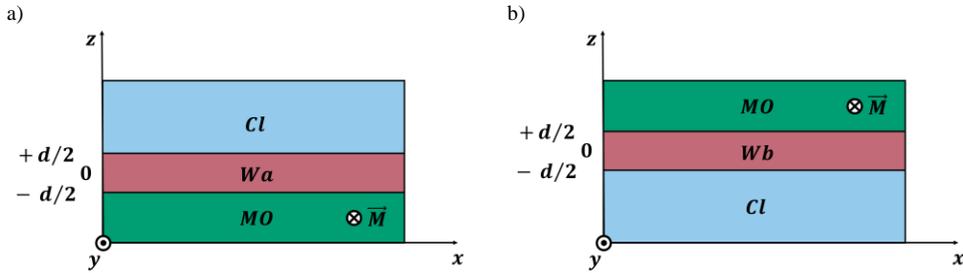

Fig. 4. Top view of the five layered heterostructure system split in two partitions as three-layered heterostructures in which the wave equations can be solved individually for waveguide a) *Wa* and b) *Wb*.

For the three-layered heterostructure depicted in Fig. 4a, the magnetic field components in *Wa* can be represented as:

$$H_y(z) = \begin{cases} A_1\exp[-K_1(z - d/2)] & z \geq d/2 \\ A_2\cos[K_2(z)] & -d/2 \leq z \leq d/2 \\ A_3\exp[K_3(z + d/2)] & z \leq -d/2 \end{cases} \quad (10)$$

Following similar steps, the dispersion relation can be expressed as follows:

$$K_2 d = \tan^{-1}\left(\frac{\gamma_{xx}^3 K_3}{\gamma_{xx}^2 K_2}\right) + \tan^{-1}\left(\frac{\beta_a \gamma_{zx}^1 + i\gamma_{xx}^1 K_1}{i\gamma_{xx}^2 K_2}\right) \quad (11)$$

$\beta_a$, represents the propagation constant of the fundamental mode of *Wa*.

For the three-layered heterostructure depicted in Fig. 4b, the magnetic field components in *Wb* can be represented as:

$$H_y(z) = \begin{cases} A_1\exp[-K_1(z - d/2)] & z \geq d/2 \\ A_2\cos[K_2(z)] & -d/2 \leq z \leq d/2 \\ A_3\exp[K_3(z + d/2)] & z \leq -d/2 \end{cases} \quad (12)$$

Following similar steps, the dispersion relation can be expressed as follows:

$$K_2 d = \tan^{-1}\left(\frac{\gamma_{xx}^1 K_1}{\gamma_{xx}^2 K_2}\right) + \tan^{-1}\left(\frac{\beta_b \gamma_{zx}^3 - i\gamma_{xx}^3 K_3}{-i\gamma_{xx}^2 K_2}\right) \quad (13)$$

$\beta_b$, represents the propagation constant of the fundamental mode of *Wb*.

The obtained relations reveal distinct propagation constants in the two scenarios, with $\beta_a \neq \beta_b$. This discrepancy arises from the inclusion of the MO layer, resulting in the disruption of both time and spatial symmetry.

To illustrate the principles we have extensively discussed, we have chosen the following values for material and geometrical parameters:

Table 1: The material and geometrical parameters used in the calculations for the propagation constants $\beta_{e,o}$ and $\beta_{a,b}$.

| Variable | Value |
| --- | --- |
| $n_{Cl}$ | 1.45 |
| $n_{Wa,b}$ | 3.48 |
| $n_{MO}$ | 1.45 |
| $g$ | 0.001, 0.005, 0.01, 0.05, 0.1 |
| $2a$ | [0.2-1.7] µm |
| $d$ | 0.3 µm |

Numerical methods can be employed to solve the dispersion relations in Eq. (8, 11 and 13) and find the propagation constants $\beta_{e,o}$ and $\beta_{a,b}$.

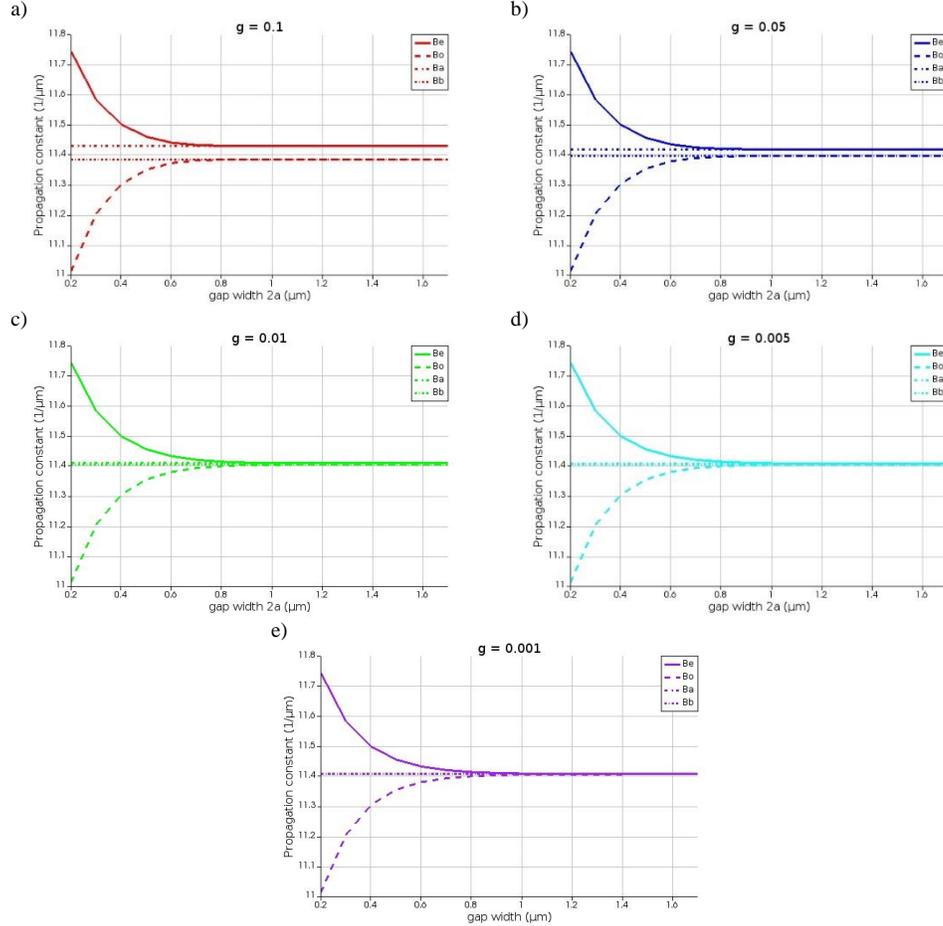

Fig. 5. Propagation constant $\beta_{e,o}$ and $\beta_{a,b}$ values for each gyrotropy case with respect to variation of MO layer width $2a$.

In all the gyrotropy cases examined above, a consistent trend emerges. For narrow gap widths ($2a$), the even and odd coupled modes exhibit strong coupling, resulting in distinct propagation constants, $\beta_e$ and $\beta_o$. However, as the width of the gap layer ($2a$) increases, $\beta_e$ and $\beta_o$ gradually approach the propagation constants of the individual waveguides $Wa$ and $Wb$, denoted as, $\beta_a$ and $\beta_b$, respectively.

## 4. Coupled Mode Theory in the waveguiding structure

The Coupled-Mode Theory (CMT) proves valuable for comprehending the coupling mechanisms within parallel waveguides. CMT assumes that the interaction between coupled modes is described by coefficients $K_{ab}$ and $K_{ba}$, which quantify the strength of the coupling. The relevant equations can be found in [18], expressed in matrix form (Eq. (14)).

$$\begin{pmatrix} P_1(x) \\ P_2(x) \end{pmatrix} = \begin{pmatrix} \cos(\psi x) - i\frac{\Delta}{\varphi}\sin(\psi x) & i\frac{K_{ab}}{\varphi}\sin(\psi x) \\ i\frac{K_{ba}}{\psi}\sin(\psi x) & \cos(\psi x) + i\frac{\Delta}{\psi}\sin(\psi x) \end{pmatrix} e^{\frac{i(\beta_a+\beta_b)}{2}x} \begin{pmatrix} P_1(0) \\ P_2(0) \end{pmatrix} \quad (14)$$

There are two eigenvalues for $\beta$: $\beta_e$ and $\beta_o$ for the even and odd modes:

$$\beta_{e,o} = \frac{\beta_a + \beta_b}{2} \pm \psi \quad (15)$$

$$\Delta = \frac{\beta_a - \beta_b}{2} \quad (16)$$

$$\psi = \sqrt{\Delta^2 + K_{ab}K_{ba}} \quad (17)$$

The coupling length which is necessary to decouple the even and odd modes can be written as a function of the propagation constants of the even and odd modes:

$$L_c = \frac{\pi}{\beta_e - \beta_o} \quad (18)$$

It is crucial to understand the propagation as well as the power exchange mechanism of the coupled modes. For this, and in the presence of MO material, we assume at the input to the parallel waveguide region: $x = 0$, and the optical power incident on both waveguides are equal, i.e., the normalized powers can be given by $P_1(0) = P_2(0) = 0.5$. The power at any point along the propagation axis can be extracted from the matrix:

$$P_1(x) = 0.5[\cos(\psi x) - i\frac{\Delta}{\psi}\sin(\psi x) + i\frac{K_{ab}}{\psi}\sin(\psi x)] e^{\frac{i(\beta_a+\beta_b)}{2}x} \quad (19)$$

$$P_2(x) = 0.5[\cos(\psi x) + i\frac{\Delta}{\psi}\sin(\psi x) + i\frac{K_{ba}}{\varphi}\sin(\psi x)] e^{\frac{i(\beta_a+\beta_b)}{2}x} \quad (20)$$

The maximum power transfer between waveguide $Wa$ and $Wb$ occurs at $L_c$, where $\psi x = \pi/2$. If we assume the power transfer between the parallel waveguides is complete, then the ratio of the normalized power difference can be written as equal to one (Eq. (20)).

$$Ratio = \frac{P_2(x)^2 - P_1(x)^2}{P_2(x)^2} = 1 \quad (21)$$

Replacing each with its equivalent term and after simplification we get:

$$Ratio = \frac{\beta_a - \beta_b}{\beta_e - \beta_o} = 0.707107 \tag{22}$$

When the $Ratio$ is 0.707107, complete power transfer occurs between the parallel waveguides. Notably, the numerator, $\beta_a - \beta_b$, remains constant as the propagation constants of the partitioned $Wa$ and $Wb$ do not depend on the MO material width ($2a$). Conversely, the denominator, $\beta_e - \beta_o$, decreases with the increase of this width leading to an increase in the quantity $Ratio$ (as well as $L_c$). Therefore, finding the width corresponding to a $Ratio = 0.707107$ is essential to identify the optimal solution.

After calculating the propagation constants $\beta_{a,b}$, $\beta_{e,o}$ through modal analysis, we can determine $L_c$ and the $Ratio$ using CMT. In Fig. 6, we have categorized the results into two sets, for $g$ values ranging between [0.1-0.01] and [0.01-0.001].

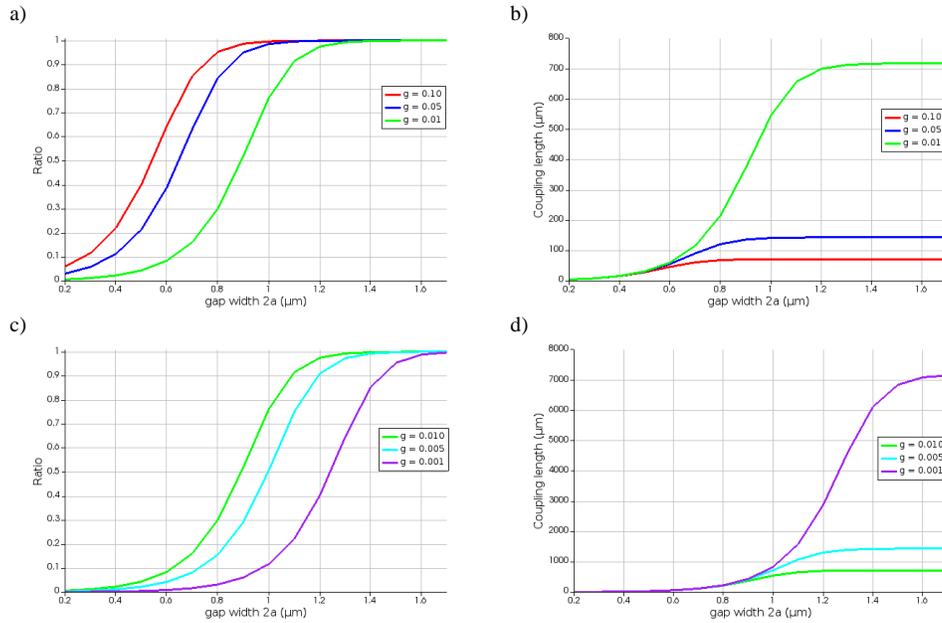

Fig. 6. (a,c) Ratio (b,d) Coupling length plots, calculated for different gyrotropy values (a,b) g = [0.1-0.01] and (c,d) g = [0.01-0.001], with respect to variation of the MO layer width $2a$.

The table below summarizes the MO gap layer width ($2a$) and the coupling length $L_c$ required in each gyrotropy case to achieve complete power transfer.

Table 2. MO layer width ($2a$) and coupling length ($L_c$) required in each gyrotropy case to achieve complete power transfer.

| Gyrotropy | $2a$ (μm) | $L_c$ (μm) |
|---|---|---|
| 0.1 | 0.62 | 50.7 |
| 0.05 | 0.73 | 101.4 |
| 0.01 | 0.97 | 506.9 |
| 0.005 | 1.08 | 1013.8 |
| 0.001 | 1.33 | 5063.5 |

As gyrotropy decreases, achieving the desired $Ratio$ requires larger gap widths ($2a$). While this may lead to an increase in coupling length, the $IL$s are confined to the waveguide material absorption, which remains below 1 dB/cm. This design presents a notable advantage: even with low gyrotropy MO material, it is feasible to create an integrable component that achieves non-reciprocal transmission with minimal $IL$s.

## 5. All dielectric isolator

For an isolator to be deemed effective, the transmission in the forward direction should be $T_f \approx 1$ (with negligible $IL$), whereas the transmission in the backward direction should be $T_b \approx 0$. Additionally, the component should not induce any back reflections $BR$ to the input ports.

$$IR = 10 \log_{10}(T_f/T_b) \quad (23)$$
$$IL = 10 \log_{10}(T_f) \quad (24)$$

At this point, although light is following different paths due the presence of the MO material, the transmission collected in both propagation directions remains equal. To address this, additional nanostructures or elements strategically placed with *Wb* are required to either absorb, diffract or redirect the backward propagating light. For instance, introducing a metal next to *Wb* can excite the plasmonic mode at the dielectric-metal interface, inducing absorption [192]. Alternatively, employing subwavelength gratings (SWGs) at the center of *Wb* with specific pitch and grating length, can operate in the radiative regime and diffract light outward [20].

Both solutions offer potential ways to significantly reduce $T_b$. In this paper, to mitigate $BR$s, we opt for a ring resonator with a high-quality factor, redirecting the light away from *Wb* [21]. The isolator component is illustrated in Fig. 7 below.

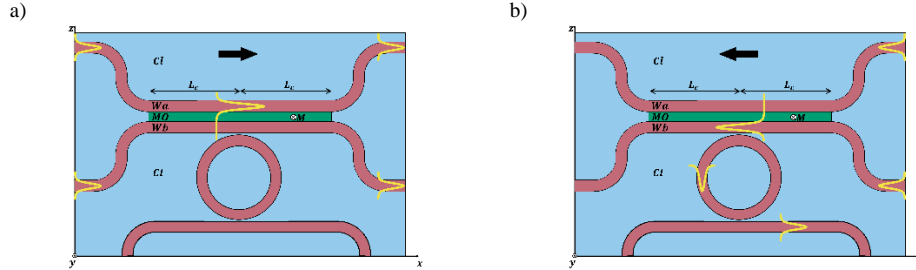

Fig. 7. Top view of the isolator component constituting the five layered heterostructure, input and output S-bends and a coupled ring resonator. Propagation in a) forward and b) backward directions.

In the forward case, following a distance of $L_c$, the coupled modes propagating simultaneously separate and the even mode appears in the middle of *Wa*. This mode remains unaffected by the ring resonator and upon traversing another $L_c$ re-excites the coupled modes. Eventually, due to presence of the S bends the coupled modes separate and reach the output ports.

In the backward case, after covering a distance of $L_c$, the coupled modes propagating simultaneously separate and the even mode emerges at the center of *Wb*. This mode is entirely diverted by the ring resonator and subsequently coupled to the drop port ensuring that no power reaches the output ports.

2D FDTD simulations were performed at $\lambda_0 = 1.55\ \mu m$, for the geometrical and material parameters presented in Table 1. While any level of gyrotropy would function in the same way, we choose the case where $g = 0.1$ to facilitate easier interpretation of the results, maintain a low computational cost, and ensure better visibility of the pictures. The ring resonator is designed with an inner radius of $6.25\ \mu m$ to provide a resonance at the desired wavelength.

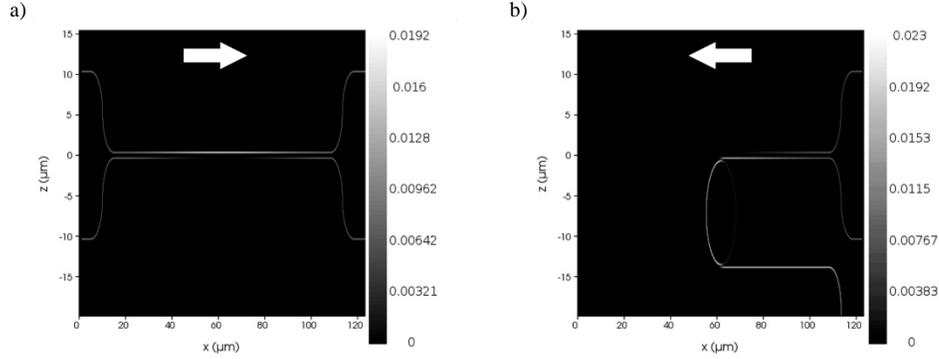

Fig. 8. Power distribution in the isolator as computed by the FDTD method (represented in greyscale). Propagation in a) forward and b) backward directions.

In the forward direction, as shown in Fig. 8a, the light from the input ports (on the left) reaches the output ports (on the right) unperturbed. The normalized transmission into the fundamental TM mode as obtained from the power monitors is $T_f = 0.98$, with an estimated $IL = -0.08\ dB$ and $BR = -22\ dB$. The marginal losses observed in the forward direction can be ascribed to attenuations resulting from the presence of S-bends in the system. In the backward direction, as shown in Fig. 8b, the light from the input ports (on the right) is redirected to the drop port and does not reach the output ports (on the left). The normalized transmission into the fundamental TM mode as obtained from power monitors $T_b = 0.006$, and $BR = -19\ dB$. Hence, the proposed isolator stands out with an $IR = 22.13\ dB$, $IL = -0.08\ dB$ and $BR = -22\ dB$.

In conclusion, simulations were conducted to ascertain the operational bandwidth by varying the wavelength from $1.54\ \mu m$ to $1.555\ \mu m$, without adjusting the MO layer width ($2a$), the coupling length ($L_c$), and the ring resonator radius. The resulting data are illustrated in Fig. 9.

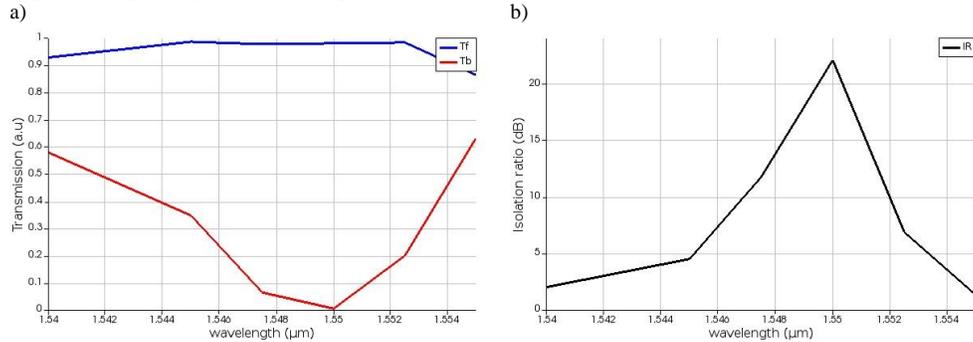

Fig. 9. a) Forward (blue) and backward (red) transmission as computed by the FDTD method. b) The calculated isolation ratio.

## 6. Conclusion

In conclusion, our study presents a thorough analysis of a dielectric isolator utilizing a five-layered heterostructure with MO material. Leveraging the CMT and a strategically introduced ring resonator, the design achieves effective non-reciprocal transmission. Notably versatile, the isolator selectively routes light paths based on external factors and exhibits high isolation, low insertion losses, and minimal back reflections.

Emphasizing tunability, the design allows for geometric adjustments to optimize performance at the desired wavelength. Furthermore, exploring SWG waveguides minimizes coupling length and decreases the footprint, offering potential for a more compact design. Such designs have also proven to enhances bandwidth. Additionally, investigating alternative nanostructures, such as those replacing the ring resonator, holds promise for further optimization, potentially increasing the operational bandwidth of the isolator even further.

These insights contribute to advancing integrated photonics, laying the groundwork for compact, reliable, and versatile optical isolators across diverse applications.

**Disclosures** The authors declare no conflicts of interest